\documentclass[twocolumn,amsmath,amssymb,prb]{revtex4}
\usepackage{graphicx}
\usepackage{dcolumn}
\usepackage{bm}
\usepackage{color}
\usepackage{ulem}

\begin{document}

\title{Tuning the Topological Features of Quantum-Dot Hydrogen and Helium by
a Magnetic Field}
\author{Wenchen Luo}
\affiliation{School of Physics and Electronics, Central South University, Changsha,
Hunan, P. R. China 410083}
\affiliation{Department of Physics and Astronomy, University of Manitoba, Winnipeg,
Canada R3T 2N2}
\author{Tapash Chakraborty}
\email{Tapash.Chakraborty@umanitoba.ca}
\affiliation{Department of Physics and Astronomy, University of Manitoba, Winnipeg,
Canada R3T 2N2}
\date{\today }

\begin{abstract}
The topological charge of the spin texture in a quantum dot with spin-orbit
couplings is shown analytically here to be stable against the ellipticity of
the dot. It is directly tunable by a single magnetic field and is related to
the \textit{sign} of the Land\'e $g$ factor. In
a quantum-dot helium, the overall winding number could have different
property from that of the single-electron case (quantum-dot hydrogen), since
tuning the number of electron affects the winding number by the Coulomb
interaction and the $z$ component angular momentum $\langle L^{}_z \rangle$.
The density profile and the spin texture influence each other when the
Coulomb interaction is present.
When $\langle L^{}_z \rangle$ is biased away from an integer by the spin-orbit
couplings, the rotational symmetry is broken which induces strong density
deformation. The sign of the topological charge may also be reversed with
increasing magnetic field. These findings are
of major significance since the applied magnetic field alone now provides a
direct route to control the topological properties of quantum dots.
\end{abstract}

\maketitle

\section{Introduction}

Studies of topological spin textures have made great strides in condensed
matter physics \cite{tokura}. The important role of spin-orbit coupling
(SOC) has been brought to the fore, in particular, since the discovery of
the topological insulator \cite{TP03,TP04}. It also plays a crucial role in
various topological states in nanoscale quantum systems. Basically, there
are two important concepts that govern the topological properties of the
states. One is that in momentum space the band structures are topologically
non-trivial, such as topological insulators and topological superconductors.
The other is that the states contain topologically non-trivial structures in
real space, such as helical magnets \cite{sky01,sky02}, skyrmions in quantum
Hall regime \cite{ezawa,bilayerg}, skyrmion lattice in non-interacting two
subband GaAs quantum well \cite{2subbandGaAs} etc. The topological band
structure is able to introduce special transport phenomena, while the
topological electron states in real space induced by the SOCs would be
important for spintroics and quantum information applications \cite{Zutic,Smejkal,Sinova}.
We have recently found \cite{lll} that the spin fields are vortices in a
quantum dot (QD) (`artificial atom') \cite{maksym,QD_book,QD01,QD03,QD04,QD05}
in the presence of SOCs \cite{QDSOC01,QDSOC02,QDSOC03,QDSOC04,QDSOC05,QDSOC06,
QDSOC07,QDSOC08,QDSOC09, QDSOC10,QDSOC11,QDSOC12,QDSOC13,QDSOC14,QDSOC15,
QDSOC16,sk,Intronati}, mostly at the single-electron level. The vortices in
real space apparently appear due to the confinement induced translational
symmetry breaking. The topological features are distinguished by the
competition of the two SOCs: the topological charge of the
spin field is $1$ when the Rashba SOC is much stronger than the Dresselhaus
SOC; and $-1$ when the Dresselhaus SOC dominates the system
in a weak magnetic field (e.g., $B\ll 1$T).

In this work, we mainly focus on how the effects of the Coulomb interaction
and a single magnetic field change the topological properties of the spin of
the electrons in quantum dots. In addition, comparison between the
many-body states and the single-electron state is necessary for that purpose.
In order to be
clearer about the topological properties of the system, we introduce the
winding number to describe the topological charge of the spin field
conveniently. So we can analytically study the spin textures with different
SOCs and magnetic fields. Particularly, we prove analytically that the
topological charge is robust against the deformation of the shape of the
quantum dot, and is tunable in a single magnetic field.

For the single electron case, by employing the perturbation theory 
{in terms of the magnetic field strength and the SOCs}, we show
that the inversion of the topological charge, when both of the Rashba and the
Dresselhaus SOCs exist, is even controllable by a single magnetic field. We
further show that the topological charge of the spin field with increasing
of the magnetic field has the opposite sign of the Land\'{e} $g$ factor,
irrespective of the single-electron or many-electron states.
We note that the perturbation theory is valid in an arbitrarily
large magnetic field, since the perturbation terms always have smaller energy scales
than that of the unperturbative one.

In order to study the interacting system, we concentrate on the two-electron
QD for simplicity, namely the QD helium \cite{QD_helium1,QD_helium2}.  {We
numerically calculate the energy spectrum and the many-particle wave functions
of the QD helium, by the method of exact diagonalization. We consider the
variation of the expectation values of the $z$ component of the angular
momentum and the spin, $\langle L^{}_z\rangle$ and $\langle \sigma^{}_z \rangle$
respectively, versus the magnetic field, in analogy to the phase transition
theory. However, as the system size is finite and no
thermodynamic limit is possible, we only mention about \it{transitions}
of these quantities. For instance,} if the
curve of $\langle L^{}_z\rangle (B)$ ($\langle \sigma^{}_z \rangle
(B) $) is not differentiable, then the discontinuous transitions occur
(akin to first-order transitions). If this curve is differentiable, but with
some clear plateaus develop, then the smooth transitions occur (similar to a
change to a second-order transition).

If there is no SOC in such a system, the spin
field would be trivial, only the $z$ component of the spin exists, and
$\langle L^{}_{z}\rangle $ is quantized with increasing
of the magnetic field. In fact,
the SOCs are able to turn the transition of $L^{}_{z}$ to second order even in
an isotropic dot, as $L^{}_{z}$ does not commute with the SOCs. The rotational
symmetry of the density may be broken when $\langle L^{}_{z} \rangle $ is no
longer an integer. The system becomes even more interesting if both spin and
density are controlled by the magnetic field or the electric field (Rashba
SOC is tuned by the electric field \cite{Rash01,Rash02,Rash03,Rash04,Rash05}%
). We thus report here how the combination of the Coulomb interaction,
Rashba and Dresselhaus SOCs modify the spin textures of these systems. Since
the density profile and the spin textures are closely dependent on and
influence each other, we also study the evolution of both density and spin with
the magnetic field.

We further explore the relations between the topological charge of the spin
field and $\langle L^{}_{z}\rangle $ or $\langle \sigma^{}_{z} \rangle $, which is
controlled by the external magnetic field. Quite remarkably, since the
topological features are directly tunable by the Land\'{e} $g$ factor and the
number of electrons in our present approach, the sign of the Land\'{e} $g$
factor which is difficult to determine experimentally \cite{gfactor} may be
addressed by detecting the topological features of the electrons properly.

The manuscript is organized as follows. In Section II, we introduce the
Hamiltonians of the QD hydrogen and the QD helium, and explicitly express the
winding number to define the topological charge of the in-plane spin
field. We write down the definition of the spin fields and the density which
will be studied in details in the following sections. Then we demonstrate
that the topological charge is robust against the ellipticity of the
dots in Section III. We also derive the topological charge to have the
opposite sign of the Land\'e $g$ factor in strong magnetic fields, which can
be verified numerically. In Section IV, we consider an InAs and a ZnO QD
helium. We study how the spin texture and the density distribution of the
electrons involves with increasing of the magnetic field. Then we study the
relation among spin fields, $\langle L^{}_z \rangle, \langle \sigma^{}_z \rangle$,
and the Coulomb interaction on the topological charge of the system
particularly. Finally, we conclude this work.

\section{The Hamiltonian and the winding number of the spin field}

The single-electron Hamiltonian in a QD with SOCs is
\begin{eqnarray}
&&\mathcal{H}=\frac{\mathbf{P}^{2}}{2m^{\ast }}+\frac{m^{\ast }}{2}\left(
\omega_x^2x^2+\omega_y^2y^2\right) +\frac{\Delta }2\sigma_z^{}+\mathcal{H}^{}_{SOC}, \\
&&\mathcal{H}^{}_{SOC}=g_1^{}\left(\sigma_x^{}P_y^{}-\sigma_y^{}P_x^{}\right)+g_2^{}\left(
\sigma_y^{}P_y^{}-\sigma_x^{}P_x^{}\right),
\end{eqnarray}%
where $\omega^{}_x$ and $\omega^{}_y$ describe the parabolic confinements in
$x$ and $y$ direction, respectively. $\sigma^{}_i$ is the Pauli matrix and
the strengths of the Rashba and Dresselhas SOCs are $g_1^{}$ and
$g_2^{}$ respectively. $P_i^{}=p_i^{}+eA_i^{}$ is the kinetic
momentum. The vector potential is chosen to be in the symmetric gauge
$\mathbf{A}=\frac12 B\left(-y,x,0\right)$ with the magnetic field $B.$
The Zeeman coupling is then $\Delta =g\mu_B^{}B$, where $g$ is the Land%
\'{e} factor. In fact, we can rewrite the Hamiltonian in the form of
\begin{eqnarray}
\mathcal{H} &=&\mathcal{H}_0^{}+\mathcal{H}_{L_z^{}}^{}+\mathcal{H}^{}_{SOC}, \\
\mathcal{H}_0^{} &=&\frac{\mathbf{p}^2}{2m^{\ast}}+\frac{m^{\ast}}2%
\left( \Omega_x^2x^2+\Omega_y^2y^2\right) +\frac{\Delta }{2}%
\sigma^{}_z, \\
\mathcal{H}_{L^{}_z} &=&\frac{eB}{2m^*}\left(
xp^{}_y-yp^{}_x\right),
\end{eqnarray}%
where $\mathcal{H}_0^{}$ describes a two-dimensional harmonic
oscillator, and $\mathcal{H}_{L^{}_z}$ is propotional to the $z$%
-component of the angular momentum $L_z^{}$. In the following
perturbative calculations, $\mathcal{H}_0$ is the unperturbated
Hamiltonian, and its
eigenvectors are chosen to be the basis in the numerical exact
diagonalizations. We introduce the frequencies $\Omega_{x,y}^{}=\sqrt{%
\omega _{x,y}^2+\omega_c^2/4}$ with the anisotropic confinement
frequencies in two directions, and the cyclotron frequency $\omega
^{}_c=eB/m^{\ast}$. The confinement lengths are $R^{}_i=\sqrt{\hbar
/(m^{\ast}\omega^{}_i)},$ and the natural length are $\ell_i^{}=%
\sqrt{\hbar /(m^{\ast }\Omega^{}_i)}$. The eigen state of the
harmonic oscillator with its eigen wavefunction $\psi^{}_{n^{}_x,n^{}_y}
(\mathbf{r}),$ where $n^{}_{x,y}$ are the
quantum numbers of the oscillator, is used as the basis of the
calculations.

If there are more than one electrons in the QD, we need to take the
Coulomb interaction into account, which can be written, in the second
quantization, \cite{lll,QD_book,QDSOC06,QDSOC09}
\begin{eqnarray}
\mathcal{H^{}_C}=\frac12\sum^{}_{i,j,k,l} \sum^{}_{s,s'} V^{}_{i,j,k,l}
c_{i,s}^{\dag }c_{j,s'}^{\dag }c^{}_{k,s'}c^{}_{l,s},
\end{eqnarray}
where $c$ is the operator of an electron, $s,s'$ are the spin indices,
$i,j,k,l$ stand for the combination indices which contains the $x,y$ quantum
numbers of the oscillator. For example, $k=\left(k^{}_x,k^{}_y\right) $,
$k^{}_{x,y}$ are the quantum numbers of the oscillator in $x$ and $y$ directions,
respectively. The Coulomb interaction matrix element $V^{}_{i,j,k,l}$ is calculated
numerically by a two-dimensional integral,
\begin{eqnarray}
&& V^{}_{ i,j,k,l} =\frac2{\pi} \frac{e^2}{\epsilon
\sqrt{\ell^{}_x\ell^{}_y}} \left( -1\right)^{\left\vert j^{}_x-k^{}_x\right\vert
+\left\vert j^{}_y-k^{}_y\right\vert } \\
&& \gamma \left( i^{}_x,l^{}_x\right) \gamma \left( i^{}_y,l^{}_y\right)\gamma
\left( j^{}_x,k^{}_x\right) \gamma \left( j^{}_y,k^{}_y\right)   \nonumber \\
&& i^{\left\vert i^{}_x-l^{}_x\right\vert +\left\vert
i^{}_y-l^{}_y\right\vert +\left\vert j^{}_x-k^{}_x\right\vert +\left\vert
j^{}_y-k^{}_y\right\vert }  \nonumber \\
&&\int dx dy\, \Phi \left( i_x,l_x,x\right) \Phi
\left( j_x,k_x,x\right)  \frac{\Phi \left( i_y,l_y,y\right)
\Phi \left( j_y,k_y,y\right) }{\sqrt{\frac{\ell _{y}}{\ell _{x}}x^{2}+\frac{
\ell _{x}}{\ell _{y}}y^{2}}}, \nonumber
\end{eqnarray}%
where $\epsilon$ is the dielectric constant and
\begin{eqnarray}
\gamma \left( n,m\right)  &=&\sqrt{\frac{2^{\min \left( n,m\right) }\min
\left( n,m\right) !}{2^{\max \left( n,m\right) }\max \left( n,m\right) !}},
\\
\Phi \left( n,m,x\right)  &=&x^{\left\vert n-m\right\vert }e^{-\frac{1}{4}%
x^{2}}L_{\min \left( n,m\right) }^{\left\vert n-m\right\vert }\left( \frac{%
x^{2}}{2}\right)
\end{eqnarray}%
with the Laguerre polynomial $L$. We note that the integrand is even ($
\left\vert i_x-l_x\right\vert + \left\vert i_y-l_y\right\vert +\left\vert
j_x-k_x\right\vert +\left\vert j_y-k_y\right\vert$ is even),
otherwise the integral would be zero, which guarantees the Coulomb
interaction matrix element to be real.

We then exactly diagonalize the total Hamiltonian $\mathcal{H_T=H+H_C}$ to
obtain the wave function of the state that we would like to study. The
selected state is supposed to be
\begin{equation}
\left\vert \Psi\right\rangle =\sum_{\left\{ j\right\} }d_{j}\left\vert
(j_{1},s_1),(j_{2},s_2),\ldots (j_{N_{e}},s_{N_e})\right\rangle ,
\end{equation}%
where $N_{e}$ is the electron number and $d_{j}$ is the coefficient of the
many-particle (or a single-particle) basis obtained by the exact
diagonalization.  { In the many-particle state $\left\vert (j_{1},s_1),
(j_{2},s_2),\ldots (j_{N_{e}},s_{N_e})\right\rangle$, $(j_{n},s_n)$ are the
indices for the $n-$th electron of the system, where $j_{n}$ is the
combination index containing the $x,y$ quantum numbers of the oscillator and
$s_n$ is the spin index. }
The spin fields $\sigma _{\mu }\left( \mathbf{r}\right) $ of
such a state can be defined generally by
\begin{eqnarray} \label{spin}
&& \sigma _{\mu }\left( \mathbf{r}\right)  =\sum_{\left\{ i\right\} ,\left\{
j\right\} }d_{i}^{\ast }d_{j}\sum_{k,l,s,s'}\psi _{k,s}^{\dag }\left( \mathbf{r}%
\right) \sigma _{\mu }\psi _{l,s'}\left( \mathbf{r}\right)    \\
&& \left\langle (i_{1},s'_1),\ldots (i_{N_{e}},s'_{N_e})\right\vert
c_{k,s}^{\dag}c_{l,s'}\left\vert (j_{1},s_1),\ldots (j_{N_{e}},s_{N_e})
\right\rangle ,   \nonumber
\end{eqnarray}%
and the density is given by
\begin{eqnarray} \label{density}
&& n\left( \mathbf{r}\right)  =\sum_{\left\{ i\right\} ,\left\{
j\right\} }d_{i}^{\ast }d_{j}\sum_{k,l,s}\psi _{k,s}^{\dag }\left( \mathbf{r}%
\right) \psi _{l,s}\left( \mathbf{r}\right)    \\
&& \left\langle (i_{1},s'_1),\ldots (i_{N_{e}},s'_{N_e})\right\vert
c_{k,s}^{\dag}c_{l,s}\left\vert (j_{1},s_1),\ldots (j_{N_{e}},s_{N_e})
\right\rangle .   \nonumber
\end{eqnarray}
The wave function of $\psi
_{k,s}\left( \mathbf{r}\right)=\psi
_{k_{x},k_{y},s}\left( \mathbf{r}\right) $ is a spinor, which can be
written explictly for different spins, $\psi _{k_{x},k_{y},+}\left( \mathbf{r%
}\right) =\left(
\begin{array}{c}
\psi _{k_{x},k_{y}}\left( \mathbf{r}\right)  \\
0
\end{array}
\right) $ and $\psi _{k_{x},k_{y},-}\left( \mathbf{r}\right) =\left(
\begin{array}{c}
0 \\
\psi _{k_{x},k_{y}}\left( \mathbf{r}\right)
\end{array}
\right) $.

In order to study the relation between the topological charge and the
environment of the QD, we explicitly define the winding number
\begin{equation}
q=\frac{1}{2\pi }\oint d\phi =\frac{1}{2\pi }\oint \frac{\sigma
_{x}^{{}}\,\left( \mathbf{r}\right) d\sigma _{y}^{{}}\left( \mathbf{r}%
\right) -\sigma _{y}^{{}}\,\left( \mathbf{r}\right) d\sigma _{x}^{{}}\left(
\mathbf{r}\right) }{\sigma _{x}\left( \mathbf{r}\right) ^{2}+\sigma
_{y}\left( \mathbf{r}\right) ^{2}},
\end{equation}%
where $\phi \left( \mathbf{r}\right) =\arctan [\sigma _{y}^{{}}\left(
\mathbf{r}\right) /\sigma _{x}^{{}}\left( \mathbf{r}\right) ].$ The route of
the integral is a closed path around a singularity of the $\phi $ field,
then we can find the winding number of this particular vortex. However, in
many-particle cases, there may be more than one vortices in the QD. So it is
also worthy defining the overall winding number (OWN) of which the path is
chosen around the edge and encloses all the possible vortices corresponding
more than one singularity points of the $\phi $ field. The winding number is
directly related to the topological feature of the in-plane spin field, and
is therefore defined as the topological charge of the system.

\section{Topological features related to the sign of the Land\'e $g$ factor}

In this work, we mainly focus on how the Coulomb interaction effects the
topology of the system. But before that we should determine some features
in the single-particle case, especially the topic how the topological charge
is varied by the magnetic field. Then the many-body effect will be clear by
comparing the two cases.

We study the generic case of an anisotropic QD in a magnetic field
perturbatively. The unperturbed ground state of $\mathcal{H}_{0}^{{}}$ is
determined by the sign of the Land\'{e} $g$ factor if $B>0$,
$\psi _{-}^{\left( 0\right) }=\left(
\begin{array}{cc}
\psi _{0,0}^{{}} & 0%
\end{array}%
\right) ^{T}$ for $g<0$ or $\psi _{+}^{\left( 0\right) }=\left(
\begin{array}{cc}
0 & \psi _{0,0}^{{}}%
\end{array}%
\right) ^{T}$ for $g>0.$ The perturbation is then $\mathcal{H}%
_{L_{z}^{{}}}^{{}}$ plus the SOCs. The wavefunctions with the
first-order corrections are
\begin{eqnarray}
\psi _{+}^{(1)} &=& \left(
	\begin{array}{c}
	\psi_{0,0} +i W \psi_{1,1}/2 \\
	\left( \Gamma_{1,x}+i \Gamma_{2,x}\right) \psi _{1,0}+\left(
    \Gamma_{2,y}^{\ast}+i\Gamma_{1,y}^{\ast }\right) \psi _{0,1}
	\end{array}
\right), \notag \\ \label{psi+} \\
\psi _{-}^{(1)} &=& \left(
	\begin{array}{c}
	\left( \Gamma_{1,x}-i\Gamma_{2,x}\right) \psi _{1,0}+\left(
    \Gamma_{2,y}^{\ast}-i\Gamma_{1,y}^{\ast }\right) \psi _{0,1} \\
	\psi_{0,0} + i W \psi_{1,1}/2
	\end{array}%
\right) , \notag \\ \label{psi-}
\end{eqnarray}
where
\begin{eqnarray}
\Gamma_{1,\left( x,y\right) } &=&\frac{1}{\sqrt{2}(|\Delta|+\hbar\Omega_{x,y})}
\left( -g_{1}\frac{eB}{2} \ell _{ x,y }+g_{2}i\frac{\hbar }
{\ell _{x,y }} \right) , \notag \\ \label{g1} \\
\Gamma_{2,\left( x,y\right) } &=&\frac{1}{\sqrt{2}(|\Delta|+\hbar\Omega_{x,y})}
\left( -g_{2}\frac{eB}{2}\ell _{ x,y }+g_{1}i\frac{\hbar }
{\ell _{x,y }}\right) . \notag \\ \label{g2} \\
W &=& \frac{\hbar \omega_c}{2\sqrt{\Omega_x \Omega_y}}\frac{\Omega_y-\Omega_x}
{-\hbar(\Omega_x+\Omega_y)}, \label{w}
\end{eqnarray}
where $W$ is the anisotropic parameter and describes the anisotropy of the dot.
Note that the wavefunctions are not normalized, but the normalization does
not change the winding number. The in-plane spin fields can be
calculated by Eq. (\ref{spin}),
\begin{eqnarray}
\sigma_{x}^{\pm }\left( \mathbf{r}\right) &=& \left[ G_{1,x}^{\pm }\frac{x}{\ell_x}+
G_{2,y}^{\pm }\frac{y}{\ell_y} \mp \left( G_{2,x}^{\pm}\frac{x}{\ell_x}
+ G_{1,y}^{\pm }\frac{y}{\ell_y} \right) W\frac{xy}{\ell_x \ell_y} \right] \notag \\
&\times& \psi_{0,0}\left( \mathbf{r}\right), \label{sx} \\
\sigma _{y}^{\pm }\left( \mathbf{r}\right) &=& \left[ G_{2,x}^{\pm}\frac{x}{\ell_x}+
G_{1,y}^{\pm }\frac{y}{\ell_y} \pm \left( G_{2,x}^{\pm}\frac{x}{\ell_x}
+ G_{1,y}^{\pm }\frac{y}{\ell_y} \right) W\frac{xy}{\ell_x \ell_y} \right] \notag \\
&\times& \psi_{0,0}\left( \mathbf{r}\right), \label{sy}
\end{eqnarray}
where
\begin{eqnarray}
G_{1,\left( x,y\right) }^{\pm }&=&\frac{\left( \pm 2\hbar -eB\ell_{x,y}^{2}
\right) g_{1} } {2 (|\Delta|+\hbar\Omega_{x,y}) \ell _{x,y} }, \\
G_{2,\left( x,y\right) }^{\pm }&=&\frac{\left( \mp 2\hbar -eB\ell _{x,y}^{2}
\right) g_{2} }{2 (|\Delta|+\hbar\Omega_{x,y}) \ell _{x,y}}.
\end{eqnarray}
Note that $\ell_{x,y}^2 = \hbar/(m^* \Omega_{x,y}) < 2 \hbar/(m^*
\omega_c) $ and $eB\ell_{x,y}^2 <2\hbar$, then we have $\mathrm{sign}
\left[G_{1,(x,y)}^\pm \right] = \pm \mathrm{sign}(g_1) $ and $\mathrm{sign}
\left[G_{2,(x,y)}^\pm \right] =\mp \mathrm{sign}(g_2) $.

Specifically, if only the Rashba SOC is present, then (see the appendix)
\begin{eqnarray} \label{rq}
q &=& \frac{G_{1,x}^\pm G_{1,y}^\pm }{\sqrt{(G_{1,y}^\pm  G_{1,x}^\pm )^{2}}}.
\end{eqnarray}%
Hence, $q=1$ whatever the sign of $g_1$ is, which means that the sign of
Rahsba SOC, or say the direction of the external electric field, does not
change the topological charge. In the same manner, if only the Dresselhaus
SOC is present, then $q=-1$. Note that the calculation is based on an
anisotropic dot and the result is not related to the anisotropic coefficient
$W$. Therefore, the topological charge is analytically proven to be robust against
the ellipticity of the dot, which has been studied only numerically in Ref.
\cite{lll}.

If both SOCs are present, the integral above becomes more complex. We
are still able to obtain that (see the appendix)
\begin{equation} \label{tc}
q=\mathtt{sgn} (G_{1,x}^{\pm}G_{1,y}^{ \pm}-G_{2,x}^{\pm}G_{2,y}^{\pm}).
\end{equation}
Consequently, we conclude that the
topological charge is still $q=\pm1$ when $G_{1,x}^{\pm}G_{1,y}^{
\pm}>G_{2,x}^{\pm}G_{2,y}^{\pm}$ or $G_{1,x}^{\pm}
G_{1,y}^{\pm}<G_{2,x}^{\pm}G_{2,y}^{\pm},$ respectively.
Note that the sign of the topological charge is not only related to the either
the strength or the sign of the SOC $g^{}_{1,2}$.
Only when the magnetic field is weak, the
strength of the SOC can determine the topological charge. But, surprisingly, it
is primarily determined by the sign of the Land\'{e} factor in a strong
magnetic field.

If we consider a strong magnetic field $B\rightarrow \infty $,
then $ eB\ell_{x,y}^2\rightarrow 2\hbar$, so that $G^+_{1,(x,y)} \rightarrow 0,
G^-_{2,(x,y)} \rightarrow 0$ and $G^-_{1,(x,y)} < 0, G^+_{2,(x,y)} < 0$.
If $g>0$, then $q \rightarrow \mathtt{sgn} \left( -G_{2,x}^+ G_{2,y}^+
\right) = -1 $, while $q \rightarrow \mathtt{sgn} \left( G_{1,x}^- G_{1,y}^-
\right) = 1 $ if $g<0$. Hence, we obtain that
\begin{equation}
q=-\mathtt{sgn}\left(g\right). \label{qlande}
\end{equation}
It is safe to consider such a limitation $B\rightarrow \infty $  {in the perturbation
calculations.
The energy scale of the unperturbative Hamiltonian $\mathcal{H}_0$ is $E_0=
\hbar \Omega_x /2+\hbar \Omega_y /2$. The energy scale of $\mathcal{H}_{L_z}$
is $E_{L_z}= \hbar \omega_c /2$, and the energy scale of $\mathcal{H}_{SOC}$
is in the same order of $E_{SOC}= \hbar g_{1,2} / \ell_{x,y} + g_{1,2} eB \ell_{x,y}/2$.
We can then compare these energy scales in the limit of $B\rightarrow \infty$.
It is obvious that $E_{L_z} < E_0$ always, since $\frac{1}{2} \omega_c <
\Omega_{x,y} = \sqrt{\omega_c ^2 /4 +\omega_{x,y}^2}$. Moreover, $E_{SOC}
=g_{1,2}\sqrt{\hbar m^* \Omega_{x,y}} \left( 1+\frac{1}{2}\frac{\omega_c}{\Omega_{x,y}}
\right) < 2 g_{1,2}\sqrt{\hbar m^* \Omega_{x,y}}$, and then $E_{SOC}
/E_0 \rightarrow 0$ when the magnetic field is very large ($\Omega_{x,y}
\rightarrow \infty$).}

 {On the other hand, in the perturbative states in Eqs. (\ref{psi+})
and (\ref{psi-}), the corrections of the unperturbative states are $\Gamma$ and
$W$ shown in Eqs.~(\ref{g1}-\ref{w}). When $B \rightarrow \infty$, the anisotropic
parameter is always $W <1$, and the perturbation of the SOCs $\mathcal{H}_{SOC}$ provides
the corrections in Eqs.~(\ref{g1}) and (\ref{g2}) also approach zero, $\Gamma_{(1,2),(x,y)}
\propto 1/\sqrt{\Omega_{x,y}} \rightarrow 0$. Indeed, when $B \rightarrow \infty$, the in-plane
spin fields vanish and the spin textures disappear. However, we can
suppose a very large magnetic field where the perturbation theory is valid
and the spin textures are still available.}

The sign of the Land\'{e} factor (a difficult problem experimentally \cite{gfactor})
may therefore be obtained by detecting the topological charge. This important property in
Eq.~(\ref{qlande}) is not only valid in the single-electron case, but also works
in the two-electron case. We shall confirm this point below in the numerical calculations.

The Land\'{e} factors of the two systems, InAs dot and ZnO dot, have opposite
signs. The unique properties of the later system \cite{jochen} have been
explored only recently \cite{falson,luo,aram,falson2,chapter}. By comparing
the InAs quantum dots, and the ZnO dots, we could determine how the Land\'{e}
factor influences the topological spin textures and the density profiles of the electrons.
For the InAs dot, the effective mass of electron is $m_{\mathrm{InAs}
}^{\ast}=0.042m^{}_e$, Land\'{e} factor $g^{}_{\mathrm{InAs}}=-14$ and
the dielectric constant $\epsilon ^{}_{\mathrm{InAs}}= 14.6$. For a ZnO dot,
the effective mass is $m_{\mathrm{ZnO}}^{\ast}=0.24m^{}_e$, Land$\acute{e}$
factor $g^{}_{\mathrm{ZnO}}=4.3$ and dielectric constant $\epsilon^{}_{%
\mathrm{ZnO}}=8.5$.

\begin{figure*}[htbp]
\includegraphics[width=12.5cm]{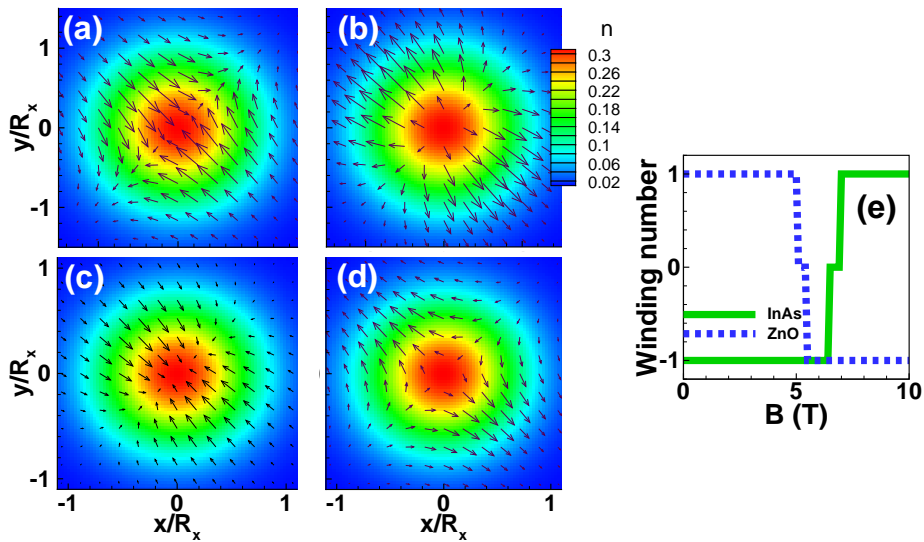}
\caption{(Colors online) The evolution of spin textures in dots with magnetic
fields. Hereafter, the colors of the two-dimensional pictures represent the
density of the electron, which is defined by Eq. (\ref{density}), and the
arrows represent the in-plane spin field $(\sigma_x (\textbf{r}), \sigma_y
(\textbf{r}) )$.
The magnetic fields are (a) $B=0.1$T, and (c) $B=10$T for an InAs dot, and (b)
$B=0.1$T, (d) $B=10$T for a ZnO dot. We consider $R^{}_x=15$ nm, $R^{}_y=14$
nm. The SOCs are $\hbar g^{}_1=\hbar g^{}_2/2=10$ meV$\cdot$ nm for the InAs
dot, and $\hbar g^{}_1=2\hbar g^{}_2=5$ meV$\cdot$ nm for the ZnO dot. (e)
The winding numbers of the two cases.}
\label{fig1}
\end{figure*}

In these single-electron systems, we consider the case when both SOCs are
present. The numerical results are shown in Fig. \ref{fig1}: with increase of
the magnetic field, we clearly see how the spin textures evolve. In the InAs
dot, if $g^{}_1>g^{}_2$, the topological charge
is always $q=1$. If $g^{}_{1}<g^{}_2$, the topological charge is $q=-1$ in a
weak magnetic field $B<6.5$T, but $q=1$ in a strong magnetic field due to
the fact that $g^{}_{\mathrm{InAs}}<0$ [Figs.~\ref{fig1}(a) and
\ref{fig1}(c)]. In a ZnO dot, if $g^{}_1<g^{}_2$, we have $q=-1$. But if
$g^{}_1>g^{}_2$, the topological charge is $q=1$ in a weak magnetic field
$B<4.5$T, while it changes to $q=-1$ in a strong magnetic field since
$g_{\mathrm{ZnO}}>0$ [Figs.~\ref {fig1}(b) and \ref{fig1}(d)]. The important
finding in Eq. (\ref{qlande}) perfectly agrees with our numerical studies
[Fig.~\ref{fig1}]. We note that the spin textures are topological trivial
$q=0$, when $G_{1,x}^{\pm }G_{1,y}^{\pm } = G_{2,x}^{\pm }G_{2,y}^{\pm}$.

\section{Quantum dot helium}

If there is more than one electron confined in the dot, we must consider the
Coulomb interaction. Indeed, the Coulomb interaction is
not negligible and provides the various magnetic signatures in the system.
We would like to determine how the Coulomb interaction affects the topological
properties of the many-electron dots. Therefore, we exactly diagonalize
$\mathcal{H}_T=\mathcal{H}+\mathcal{H}_C$ to obtain
the electron density, the spin textures, $\langle L^{}_z\rangle$, as well as
the winding number. The spin textures depend on the density
profile, and conversely, the density profile can be modified by the spin
textures. The density profile of the ground state is closely related to
$\langle L^{}_z \rangle$. We study the relations among those quantities in this
section. For simplicity and without loss of generality, we consider only the
two-electron case, viz. the quantum dot helium.

In our exact diagonalization scheme, we keep the quantum number of the harmonic
oscillator $n^{}_x,n^{}_y \in [0,5]$,  {and so that $72$ single-electron
states are taken into account.} This cut-off is sufficient for the energy
convergence.

\subsection{Two-electron states in isotropic dots with one type of SOC}

We consider the isotropic QD in this subsection.
In such a QD without the SOCs, $\langle L_z \rangle$ is quantized
with the increase of
the magnetic field. $\langle L^{}_z \rangle$ is always an integer and the
system has the rotational symmetry.

\begin{figure}[tbp]
\centering
\includegraphics[width=8 cm]{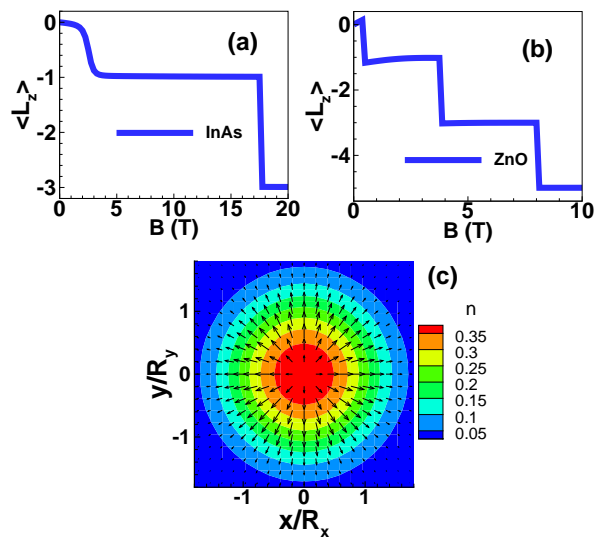}\newline
\caption{(Colors online) $\langle L^{}_z \rangle$ of the two-electron states in (a) an InAs dot
and (b) in a ZnO dot with the Rashba SOC. The size of the two dots is the same,
$R^{}_x=R^{}_y=15$ nm. The strengths of the SOC are $\hbar g^{}_1= \hbar
g^{}_2=20$ meV$\cdot$nm for the InAs dot, and $\hbar g^{}_1= \hbar g^{}_2 =5$
meV$\cdot$ nm for the ZnO dot. (c) The density and the spin texture of the
two-electron InAs dot at $B=2.5$T, where $\langle L^{}_z \rangle =-0.4867$.
}
\label{fig2}
\end{figure}

With either one of the SOCs (Rashba or Dresselhaus), $\langle L^{}_z \rangle$
is not a good quantum number, and $\langle L^{}_z \rangle$ is no longer an
integer, as shown in Figs. \ref{fig2}(a) and \ref{fig2}(b). The spin texture
is a single vortex with the topological charge $q= \pm 1$ in a magnetic field
for a Rashba dot or a Dresselhaus dot, respectively. The density profile is
still rotationally invariant. The rotation matrices for the Rashba SOC ($U^{}_R$)
and the Dresselhaus SOC ($U^{}_D$) are \cite{lll}
\begin{eqnarray}
	U^{}_R(\theta)
	=
	\begin{pmatrix}
      		\cos{\theta}&  \sin{\theta}  \\
      		-\sin{\theta}&   \cos{\theta}
	\end{pmatrix},
	\ U^{}_D(\theta)	= U^{}_R(-\theta),
\end{eqnarray}
where $\theta$ is the angle in the polar coordinate of the $x-y$ plane, are
still valid to protect the rotational symmetry of the system as well as
the density of electrons if there is only one SOC present. We show the $\langle L^{}_z \rangle $
in terms of the magnetic field in an InAs QD and in a ZnO dot with Rashba
SOC in Fig.~\ref{fig2}(a) and Fig.~\ref{fig2}(b), respectively. The density
profile and spin texture of the InAs dot are shown when $B=2.5$T and
$\langle L^{}_z \rangle =-0.4867$ in Fig. \ref{fig2}(c).

\subsection{Two-electron states in isotropic InAs dots with both two SOCs}

\begin{figure}[tbp]
\centering
\includegraphics[width=8 cm]{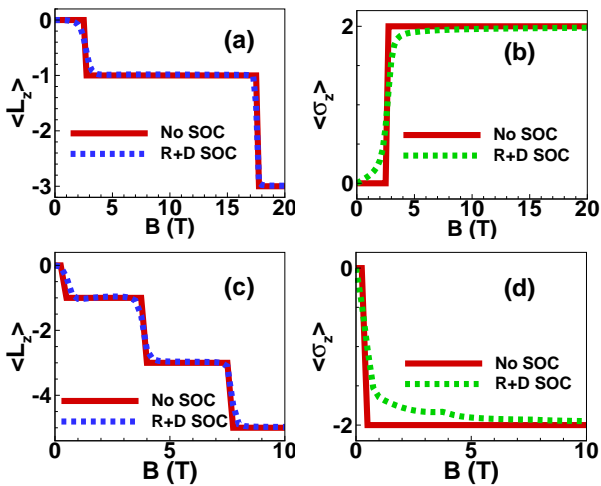}\newline
\caption{(Colors online) (a) $\langle L^{}_z \rangle$ and (b) $\langle \protect\sigma^{}_z
\rangle$ of the two-electron state in an InAs dot with and without the SOCs.
The strengths of the SOCs are $\hbar g^{}_1= \hbar g^{}_2=20$ meV$\cdot$nm.
(c) $\langle L^{}_z \rangle$ and (d) $\langle \protect\sigma^{}_z \rangle$
of the two-electron state in a ZnO dot with and without the SOCs. The
strengths of the SOCs are $\hbar g^{}_1= \hbar g^{}_2 =5$ meV$\cdot$ nm. The
size of the two dots is the same, $R^{}_x=R^{}_y=15$ nm.}
\label{fig3}
\end{figure}

$\langle L^{}_z \rangle$ displays a smooth transition with the magnetic
field when both the SOCs are present. Then the  {effective}
rotational symmetry of the density can be broken, since the spin field no
longer has the rotational symmetry. It is therefore more interesting to
discuss in detail how the two SOCs and the Coulomb interaction jointly
influence the spin textures and then the density profiles with increasing
magnetic field. For simplicity, we consider the case $ g^{}_1= g^{}_2$ in an
isotropic dot only [Fig.~\ref{fig3}]. The discontinuous transitions in
$\langle L^{}_z \rangle$ are smoothed out
in the presence of the SOCs as well. Strictly speaking, there is no integer plateau
of $\langle L^{}_z \rangle$. However, we can still mark the plateaus where
the system shows $\langle L^{}_z \rangle$ very close to being integers. We
also show $\langle \sigma^{}_z \rangle$ in Fig.~\ref{fig3}, since it is
measurable in an NMR experiment \cite{sean}.

In the InAs dot without the SOC, no spin textures appear and the density
profile evolves from a dot to a ring as the magnetic field increases, since
$ \langle L^{}_z \rangle$ jumps from $0$ to $-3$ when the magnetic field
increases up to $20$T.  { Due to the existence of the confinement, the degeneracy
of the Landau level is lifted, and the Coulomb interaction mixes
single-electron levels with different angular momentum to quantize $\langle
L_z \rangle$.} When the SOCs are present, spin textures appear to deform the
density profile, especially when $\langle L^{}_z \rangle$ is in the region
between two plateaus.

\begin{figure*}[tbp]
\centering
\includegraphics[width=16.5 cm]{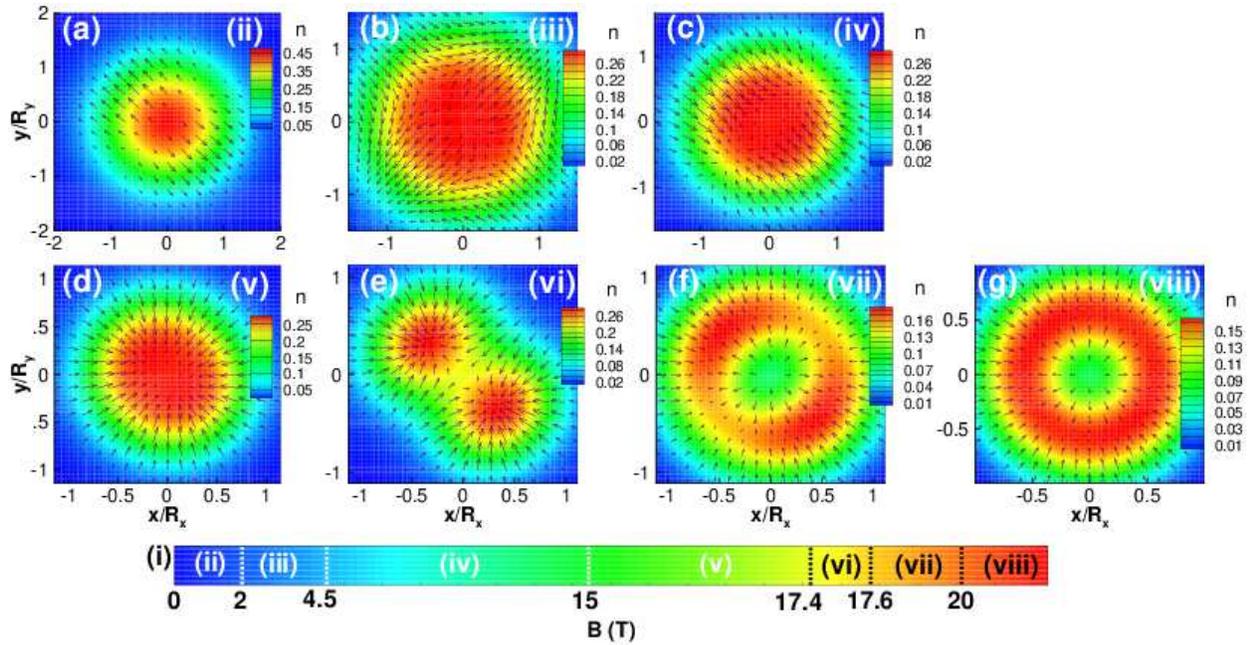}\newline
\caption{(Color online) The density profiles of a two-electron InAs dot, $
R^{}_x=R^{}_y=15$ nm, with SOCs $\hbar g^{}_1= \hbar g^{}_2=20$ nm$\cdot$
meV. The colors stands for the density $n$ of the two electrons. The magnetic
fields are (a) $1$T; (b) $3.7$T; (c) $6$T; (d) $12$T; (e) $17.5$T; (f) $18$T; (g) $23$T.
The Roman numerals are corresponding to the states described in the text.
}
\label{fig4}
\end{figure*}

Since the spin textures are much richer in quantum-dot helium than those in
a single-electron dot (QD hydrogen), we need to use the concept of the overall
winding number (OWN) which is equivalent to the total topological charge. It
can also be obtained by summing the topological charge for each vortex in the
system.

We consider here the dot with $R^{}_x=R^{}_y =15$ nm, and SOCs $\hbar g_1=
\hbar g^{}_2=20$ nm$\cdot$ meV. With an increase of the magnetic field, the
density profile and spin textures evolve as follows  {(or see the rainbow bar
in Fig. \ref{fig4}): }

(i) $B=0$T, the spin textures are cancelled by the time reversal symmetry;
(ii) $0<B<2$T, the density profile and spin textures are close to the
single-particle case shown in Fig.~\ref{fig4}(a). $\langle L^{}_z \rangle$ is
in the first plateau $\langle L^{}_z \rangle \approx 0$;
(iii) $2<B<4.5$T, the spin textures are shown in Fig.~\ref{fig4}(b). There
are three vortices located along the line $x=-y$, in which one with $q=1$
locates at the center and the other two with topological charge $q=-1$
locate at $(-0.4,0.4)$ and $(0.4,-0.4)$. The OWN is thus obtained by summing all
the three charges, $q=-1$. The density is
an elliptic dot stretched along $x=-y$, since $\langle L^{}_z \rangle$ is
between the two plateaus;
(iv) $4.5<B<15$T, the spin textures are shown in Fig.~\ref{fig4}(c), there are
still three vortices: two have topological charge $q=1$ [locate at (0.5,0.5)
and (-0.5, -0.5)] and the other has
$q=-1$ at the origin, but in the line of $x=y$. The OWN is changed to $+1$. The density
is again close to that of an isotropic dot, since $\langle L^{}_z \rangle$ is
located on the plateau;
(v)$15<B<17.3$T, the three vortices merge toward the center, while the density
is again stretched but along the line $x=-y$, which is shown in Fig. \ref{fig4}(d);
(vi) Magnetic field around $17.5$T, where $\langle L^{}_z \rangle$ is between
two plateaus $-1$ and $-3$, the density is split into two dots [Fig. \ref{fig4}(e)],
i.e., significantly different from all the other cases. Two vortices with
$q=1$ locate at the two density dots, and one with $q=-1$ at origin. The OWN
is thus $1$;
(vii) $17.7<B<20$T, the split rings of density is shown in Fig.~\ref{fig4}(f);
(viii) $B>20$T, the density is not split by the SOCs [Fig.~\ref{fig4}(g)],
since $\langle L^{}_z \rangle \approx -3$ is again back to the plateau, and
the OWN is still $1$.

\subsection{Two-electron states in isotropic ZnO dots with both two SOCs}

We now consider the ZnO dot where the Coulomb interaction is much stronger
than that for the InAs dot \cite{aram}. Without the SOC, the density has a
ring-shape even at $\langle L^{}_z \rangle=0$. With the SOCs we are able to
study how important the role of Coulomb interaction is in splitting the
density. Note that the SOCs in ZnO is much weaker, so we consider $\hbar
g^{}_1= \hbar g^{}_2=5$ meV$\cdot$ nm.

\begin{figure*}[tbp]
\centering
\includegraphics[width=16.5cm]{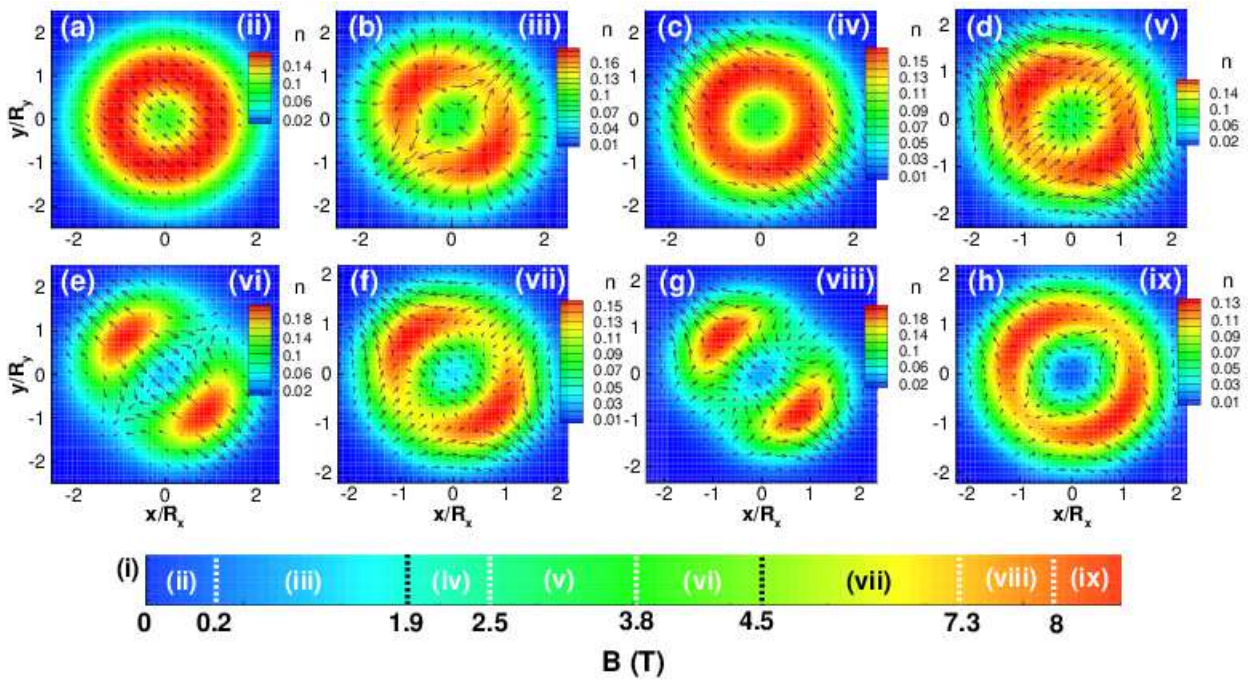}
\caption{(Color online) The density profiles of a two-electron dot, $%
R^{}_x=R^{}_y =15$ nm, with SOCs $\hbar g^{}_1= \hbar g^{}_2=5$ nm$\cdot$
meV. The magnetic fields are (a) $0.1$T; (b) $0.7$T; (c) $2.2$T; (d) $3$T;
(e) $4$T; (f) $7$T; (g) $7.5$T; (h) $9$T.}
\label{fig5}
\end{figure*}

Just as in the case of the InAs dot, we find the following spin-density
textures with increase of the magnetic field up to $10$T  {(or see the rainbow
bar in Fig. \ref{fig4}):}

(i) $B=0$, the spin textures are cancelled by the time reversal symmetry;
(ii) $0<B<0.2$T, the spin textures and the density are close to that of the
single-particle case [Fig.~\ref{fig5}(a)];
(iii) $0.2<B<1.9$T [Fig. \ref{fig5}(b)]: $\langle L^{}_z \rangle$ is between
the two plateaus $\langle L^{}_z \rangle \approx 0$ and $\langle L^{}_z
\rangle \approx -1$. So the density splits and meanwhile the spin textures
split, two vortices with $q=1$ at (-0.8, 0.8) and (0.8,-0.8) and one vortex
with $q=-1$ at origin. The OWN is then $1$;
(iv) $1.9<B<2.5$T [Fig.~\ref{fig5}(c)], the OWN changes to $q=-\mathtt{sgn}
\left( g^{}_{\mathrm{ZnO}} \right)=-1$. The reason is the same as for the
InAs dot -- the single-electron treatment;
(v) $2.5<B<3.8$T [Fig.~\ref{fig5}(d)], the density starts to split between
the two plateaus $\langle L^{}_z \rangle \approx -1$ and $\langle L^{}_z
\rangle \approx -3$. The density profile is similar to the case (iii) but
the topological features are inversed, two vortices with $q=-1$ at (-0.6,
0.6) and (0.6, -0.6) while one vortex with $q=1$ at origin. The OWN is $-1$;
(vi) $3.8<B<4.5$T [Fig.~\ref{fig5}(e)], the density profile has a two-dot
shape. The spin textures are not very regular, but we can find that two
vortices at the two dots are with topological charge $1$ and the third one
is $q=-1$ at origin;
(vii) $4.5<B<7.3$T [Fig. \ref{fig5}(f)], the density merges to a split ring,
when $\langle L^{}_z \rangle$ enters to a plateau. The spin textures are
even more complex, three vortices with $q=-1$ at (-0.8,0.8), (0,0), and
(0.8,-0.8) and two vortices with $q=1$ at (0.5,0.5) and (-0.5,-0.5), so the
OWN is $q=-1$;
(viii) $7.3<B<8$T [Fig. \ref{fig5}(g)], the density splits to two dots again
between the two plateau between the two plateaus $\langle L^{}_z \rangle
\approx -3$ and $\langle L^{}_z \rangle \approx -5$. The two vortices with
$q=-1$ are associated with the two density dots and one vortex with $q=1$ at
origin;
(ix) $B>8$T [Fig. \ref{fig5}(f)], the density merges to a split ring, again.
The OWN is $-1$, however, the details of the vortices in the ring are not
very clear in the figure.

\subsection{Summary of the density evolution in the isotropic
QD helium with both SOCs included}

Overall, the density profile evolves with the change of $\langle L^{}_z
\rangle$. Generally for the density profile, whatever the material is, the
dot-shape is stretched and the ring-shape splits by the SOCs when $\langle
L^{}_z \rangle$ is far away from an integer, while it merges when $\langle
L^{}_z \rangle$ enters a plateau near an integer. The evolution of the
density profile with increase of magnetic field falls to a split-merge cycle.

\subsection{Coulomb interaction effects}

The deformation of the density profile can be understood as following. When
$\langle L^{}_z \rangle$ is in the plateau, the many-body state is basically
composed by the eigen states of $L_z$, $|\Psi \rangle \approx |n_1,l_1,s_1
\rangle |n_2, l_2,s_2 \rangle$, where $l_1,l_2$ are the quantum number of
$L_z$ and $l_1+l_2= \langle L^{}_z \rangle$, $n_1,n_2$ are Landau level
indices and $s_1,s_2$ are the spin indices of the system.
The density has the rotational symmetry since $\langle L^{}_z \rangle$ is still
an integer and $L_z$ commutes with rotation with respect to the $z$ axis.
The SOCs accompanied by the Coulomb interaction make the $\langle L^{}_z \rangle$
deviate from the plateaus. Then the wave function may not be so much related
to two eigenstates of $L_z$ with the same phase (the wave function is given
by the superposition of many eigenstates of $L_z$ with complex phases).
Moreover, both of the two SOCs exist and neither $U_R$ nor $U_D$
protects the rotational symmetry, since $L_z \pm \sigma_z/2$ does not commutate
with the SOCs any more. So the rotational symmetry of the wave
function and the density can be broken. On the other hand, the density
deformation induces the changes of the Coulomb interaction. When all these
conditions combine together, the density splitting occurs only when $\langle
L^{}_z \rangle$ is far away from the integer plateau.  {If there is no Coulomb
interaction, then no mixing among single-electron levels appears}, so that
$\langle L^{}_z \rangle \rightarrow \sim -1$ and no more
transition happens, then the split-merge cycle of the density deformation
disappears. The non-interacting picture is apparently incorrect in the
many-particle cases.

\begin{figure*}[tbp]
\centering
\includegraphics[width=16.cm]{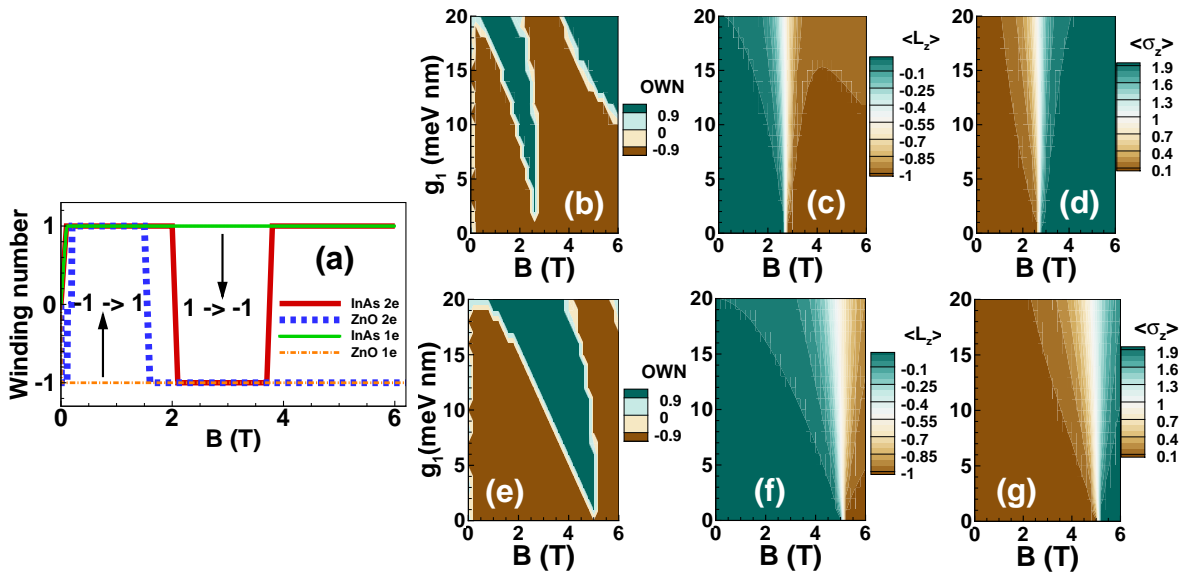}
\caption{(Color online) (a) The OWN of the QD helium where $R_x=R_y =15$ nm,
$\hbar g^{}_1= \hbar g^{}_2=20$ nm$\cdot$ meV for the InAs dot, and
$R^{}_x=R^{}_y =15$ nm, $\hbar g^{}_1= \hbar g^{}_2=5$ nm$\cdot$
meV for the ZnO dot. For comparison the
OWNs of the QD hydrogen with $g_1=g_2$ are also shown. The arrows show the
OWN reversed by the Coulomb interaction. (b) The OWN, (c) $\langle L_z \rangle$, and
(d) $\langle \sigma_z \rangle$ for the InAs QD helium with tunable $\hbar g_1
\in [0,20]$meV$\cdot$nm and fixed $\hbar g_2 =20 $meV$\cdot$nm. In
comparison, (e) the OWN, (f) $\langle L_z \rangle$, and (g) $\langle \sigma_z
\rangle$ of the ground state of the two non-interating electrons are also
plotted.
 }
\label{fig6}
\end{figure*}

We note that the topological features are tunable by adding electron
in the system. We consider an InAs dot with $\hbar g_1=\hbar g_2=20$meV$\cdot$
nm. The sign of the OWN is altered in the QD helium by the many-body effect as
shown in Fig. \ref{fig6}(a). It is very interesting that the OWN is even
related to $\langle L^{}_z \rangle$. The windows of inverse of the OWNs for
the QD helium (comparing with the cases of QD hydrogen) are open when
$\langle L^{}_z \rangle$ is converted from $0$ to
$-1$ and $\langle \sigma_z \rangle$ is converted from $0$ to $ 2$. When
$B<2.8$T, the topological inversion is also related to the region $0<\langle
\sigma_z \rangle<1$ shown in Fig. \ref{fig6}(d), which provides the clue of the
indirect measurement of the topological charge.

In order to determine how the Coulomb interaction affects the topological
properties of the system, we compare the interacting two-electron state and the
non-interacting two-electron state. We consider the fixed $\hbar g_2=20$meV
$\cdot$nm and varied $g_1$ in the InAs dot. The OWNs, $\langle L_z \rangle$,
and $\langle \sigma_z \rangle$ of the two interacting electrons states are
indicated in Figs. \ref{fig6}(b), \ref{fig6}(c), and \ref{fig6}(d),
respectively. The region of the inverse of the topological charge, comparing
with the single electron state is basically covered
by the region of the transition $\langle L^{}_z \rangle = 0 \rightarrow -1$.
As discussed above, the SOCs associated with the Coulomb interaction change
the wave function mostly in this region between two plateaus of $\langle L_z
\rangle$. Hence, it is mostly possible to change the topological features in
such regions.
The OWNs, $\langle L_z \rangle$, and $\langle
\sigma_z \rangle$ of the two non-interacting electrons are shown in Figs.
\ref{fig6}(g), \ref{fig6}(h), and \ref{fig6}(i), respectively. The Coulomb
interaction significantly shifts and compresses the regions of all of these
quantities (OWNs, $\langle L_z \rangle$, and $\langle \sigma_z \rangle$).

We further note that when the magnetic field is strong, the OWN should then
be given by the single-particle case.
The contour of calculating the OWN is far away from the center of the
quantum dot, where the density of electrons is small and the Coulomb interaction
plays a less important role. The SOCs determine the topological feature
of the system, i.e. $q \rightarrow -\mathtt{sgn}\left( g \right)$.

 \subsection{Two-electron states in anisotropic dots}

We also study the more practical case: how the ellipticity affects the
spin-density profiles. Firstly, we consider a slightly strained dot, $%
R^{}_x=15$ nm and $R^{}_y=14.9$ nm. All the states in Figs.~\ref{fig4} and %
\ref{fig5} are unaltered. However, the spin textures are slightly twisted
and the density profile is rotated anticlockwise toward the $-x$ axis.

\begin{figure*}[tbp]
\centering
\includegraphics[width=16.cm]{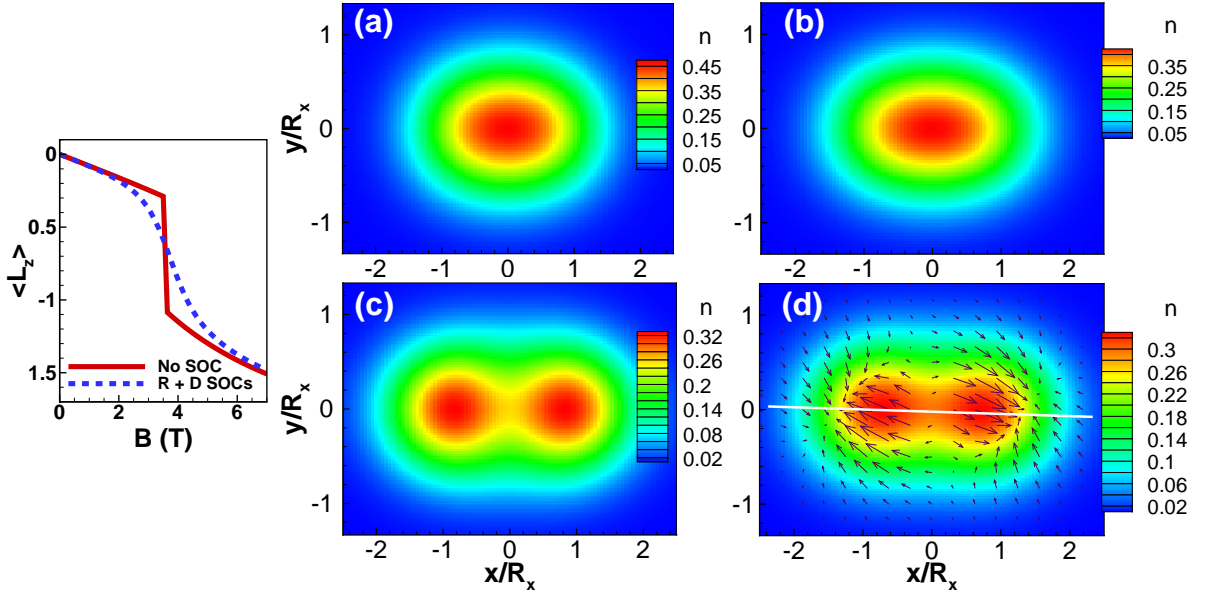}
\caption{(Color online) (a) $\langle L_z \rangle$ for the InAs quantum dot
helium $R_x= 15 $nm, $R_y = 10$nm, with and without SOCs. The SOCs are
$g_1 = g_2 =20$nm $\cdot$ meV. The density profiles of the quantum dot
without SOC at (a) $B = 0$T, (b) $B = 3$T and (c) $B = 5$T. (d) The SOCs
are coupled to such a dot at B = 5T. The white line is a guidance to eyes.
}
\label{fig7}
\end{figure*}

In Fig \ref{fig7}, the dot is strained seriously and without the SOC, e.g.
$R^{}_x=15$ nm, $R^{}_y=10$ nm. The density profile split into two dots along
the long-axis when $B>4$T, where the plateau of $\langle L_z \rangle$
disappears and $\langle L_z \rangle$ is far away from an integer.
The spin textures with only one SOC were reported earlier in \cite{lll},
the density can not be rotated. With both the SOCs, the spin textures rotate
the density clockwise slightly, $< 5^\circ$.

\section{Conclusion}

In summary, we have shown that the winding number (topological charge)
uniquely depends on the sign of Land\'{e} $g$ factor of the material in a
strong magnetic field, and in the presence of both Rashba and dresselhaus
SOCs. We also analytically demonstrate that this number is robust against
the ellipticity of the dot. With both the SOCs present, the spin textures
can deform the density profile of the quantum-dot helium, since the
transition of $\langle L^{}_z \rangle$ is smoothed.
Between two
plateaus of $\langle L^{}_z \rangle$, the rotational symmetry is broken. In
such regions, the dot-shaped density becomes stretched, while a ring-shaped
density is split, with the coupling of SOCs. The topological features at the
edge where the contour of calculating the OWN is far away from
the center follows the rule of the QD hydrogen, $q=-
\mathtt{sgn} \left(\Delta \right)$ when  {the magnetic field is sufficiently strong.}

The Coulomb interaction  {accompnied with both the SOCs} can
reverse the sign of the total topological charge
around the region of the magnetic field where $\langle L^{}_z \rangle : 0
\rightarrow -1$. A stronger Coulomb interaction can make this topological
transition happen in a weaker magnetic field. Note that we
consider only two electrons in the system, more electrons and more complex
Coulomb interaction may change the topology in a more significant way. It
perhaps indicates that in other topological non-trivial systems, the Coulomb interaction
may be also important and needs to be carefully treated.

The significance of these findings is that the topological charge of the
electron state is easily tunable by the perpendicular magnetic field alone
if both of the SOCs are intrinsic, which thus provides
to control the topology of the system in spintronics and quantum information.

\section{Acknowledgement}

W.L. acknowledges support by the NSF-China under Grant No. 11804396. W.L.
also thanks Jian Sun and Yu Zhou for helpful discussions.
Computation time was provided by Calcul Qu\'{e}bec and Compute Canada.

\begin{widetext}
\section{Appendix}

We derive the topological charge explicitly shown in Eqs. (\ref{rq}) and (%
\ref{tc}) here. We use the short note $G^{\pm }=G$ in Eqs. (\ref{sx}) and (%
\ref{sy}), and then the topological charge for a Rashba dot ($g_{1}\neq
0,g_{2}=0$) is
\begin{eqnarray*}
q &=&\frac{1}{2\pi }\oint \frac{\sigma _{x}^{R}d\sigma _{y}^{R}-\sigma
_{y}^{R}d\sigma _{x}^{R}}{\sigma _{x}^{2}+\sigma _{y}^{2}} \\
&=&\frac{1}{2\pi }\oint \left[ \left( G_{1,x}\cos \theta +r^{2}G_{1,y}W\cos
\theta \sin ^{2}\theta \right) d\left( G_{1,y}\sin \theta -r^{2}G_{1,x}W\cos
^{2}\theta \sin \theta \right) \right.  \\
&&-\left. \left( G_{1,y}\sin \theta -r^{2}G_{1,x}W\cos ^{2}\theta \sin
\theta \right) d\left( G_{1,x}\cos \theta +r^{2}G_{1,y}W\cos \theta \sin
^{2}\theta \right) \right]  \\
&&/\left[ \left( G_{1,x}\cos \theta +r^{2}G_{1,y}W\cos \theta \sin
^{2}\theta \right) ^{2}+\left( G_{1,y}\sin \theta -r^{2}G_{1,x}W\cos
^{2}\theta \sin \theta \right) ^{2}\right] ,
\end{eqnarray*}%
where $\sigma _{i}^{R}$ is the spin field with $g_{2}=0$ in Eqs. (\ref{sx})
and (\ref{sy}), and $r=\sqrt{x^{2}+y^{2}}$. Then the topological charge is
\begin{equation*}
q=\frac{1}{2\pi }\oint \frac{G_{1,x}G_{1,y}\left( 1+r^{4}W^{2}\cos
^{2}\theta \sin ^{2}\theta \right) -r^{2}W\left( G_{1,x}^{2}\cos ^{2}\theta
+G_{1,y}^{2}\sin ^{2}\theta \right) \left( \cos ^{2}\theta -\sin ^{2}\theta
\right) }{\left( 1+r^{4}W^{2}\cos ^{2}\theta \sin ^{2}\theta \right) \left(
G_{1,y}^{2}\sin ^{2}\theta +G_{1,x}^{2}\cos ^{2}\theta \right) },
\end{equation*}%
The integral is obtained by considering a circular contour of which the
center is at the origin and the radius is $r$,
\begin{eqnarray*}
q &=&\frac{1}{2\pi }\int_{0}^{2\pi }d\theta \frac{G_{1,x}G_{1,y}}{%
G_{1,y}^{2}\sin ^{2}\theta +G_{1,x}^{2}\cos ^{2}\theta }-\frac{1}{2\pi }%
\int_{0}^{2\pi }d\theta \frac{r^{2}W\cos \left( 2\theta \right) }{\left(
1+r^{4}W^{2}\cos ^{2}\theta \sin ^{2}\theta \right) }
=\frac{G_{1,x}G_{1,y}}{\sqrt{G_{1,y}^{2}}\sqrt{G_{1,x}^{2}}},
\end{eqnarray*}%
which is the same as Eq. (\ref{rq}). It is clear that the integral with $W$
in the integrand vanishes, so the elliptical effect vanishes.

For the Dresselhaus dot ($g_{1}=0,g_{2}\neq 0$), it is in the same way to
obtain that $q=-1$. The charge is
\begin{eqnarray*}
q &=&\frac{1}{2\pi }\oint \frac{\sigma _{x}^{D}d\sigma _{y}^{D}-\sigma
_{y}^{D}d\sigma _{x}^{D}}{\sigma _{x}^{2}+\sigma _{y}^{2}} \\
&=&\frac{1}{2\pi }\oint \left[ \left( G_{2,y}\sin \theta +r^{2}G_{2,x}W\sin
\theta \cos ^{2}\theta \right) d\left( G_{2,x}\cos \theta -r^{2}G_{2,y}W\cos
\theta \sin ^{2}\theta \right) \right.  \\
&&\left. -\left( G_{2,x}\cos \theta -r^{2}G_{2,y}W\cos \theta \sin
^{2}\theta \right) d\left( G_{2,y}\sin \theta +r^{2}G_{2,x}W\sin \theta \cos
^{2}\theta \right) \right]  \\
&&/\left[ \left( G_{2,y}\sin \theta +r^{2}G_{2,x}W\sin \theta \cos
^{2}\theta \right) ^{2}+\left( G_{2,x}\cos \theta -r^{2}G_{2,y}W\cos \theta
\sin ^{2}\theta \right) ^{2}\right] ,
\end{eqnarray*}%
where $\sigma _{i}^{D}$ is the spin field with $g_{1}=0$ in Eqs. (\ref{sx})
and (\ref{sy}). Then
\begin{eqnarray*}
q
&=&\frac{1}{2\pi }\oint \frac{-G_{2,y}G_{2,x}\left( 1+r^{4}W^{2}\cos
^{2}\theta \sin ^{2}\theta \right) -r^{2}W\left( G_{2,x}^{2}\cos ^{2}\theta
+G_{1,y}^{2}\sin ^{2}\theta \right) \left( \cos ^{2}\theta -\sin ^{2}\theta
\right) }{\left( 1+r^{4}W^{2}\cos ^{2}\theta \sin ^{2}\theta \right) \left(
G_{2,x}^{2}\cos ^{2}\theta +G_{2,y}^{2}\sin ^{2}\theta \right) } \\
&=&-\frac{1}{2\pi }\int_{0}^{2\pi }d\theta \frac{G_{2,y}G_{2,x}}{%
G_{2,x}^{2}\cos ^{2}\theta +G_{2,y}^{2}\sin ^{2}\theta }-\frac{1}{2\pi }%
\int_{0}^{2\pi }d\theta \frac{r^{2}W\cos \left( 2\theta \right) }{\left(
1+r^{4}W^{2}\cos ^{2}\theta \sin ^{2}\theta \right) }=-\frac{G_{2,x}G_{2,y}}{%
\sqrt{G_{2,y}^{2}}\sqrt{G_{2,x}^{2}}}.
\end{eqnarray*}

The general case with both of the two SOCs reads,
\begin{equation*}
q=\frac{1}{2\pi }\oint \frac{\sigma _{x}d\sigma _{y}-\sigma _{y}d\sigma _{x}%
}{\sigma _{x}^{2}+\sigma _{y}^{2}}=\frac{1}{2\pi }\oint \frac{\left( \sigma
_{x}^{R}+\sigma _{x}^{D}\right) d\left( \sigma _{y}^{R}+\sigma
_{y}^{D}\right) -\left( \sigma _{y}^{R}+\sigma _{y}^{D}\right) _{y}d\left(
\sigma _{x}^{R}+\sigma _{x}^{D}\right) }{\left( \sigma _{x}^{R}+\sigma
_{x}^{D}\right) ^{2}+\left( \sigma _{y}^{R}+\sigma _{y}^{D}\right) ^{2}}.
\end{equation*}%
The denominator is
\begin{eqnarray*}
&&\left( \sigma _{x}^{R}\right) ^{2}+\left( \sigma _{y}^{R}\right)
^{2}+\left( \sigma _{x}^{D}\right) ^{2}+\left( \sigma _{y}^{D}\right)
^{2}+2\sigma _{x}^{R}\sigma _{x}^{D}+2\sigma _{y}^{R}\sigma _{y}^{D} \\
&=&\left( 1+r^{4}W^{2}\cos ^{2}\theta \sin ^{2}\theta \right) \left[ \left(
G_{1,x}\cos \theta +G_{2,y}\sin \theta \right) ^{2}+\left( G_{1,y}\sin
\theta +G_{2,x}\cos \theta \right) ^{2}\right] ,
\end{eqnarray*}%
where
\begin{equation*}
2\sigma _{x}^{R}\sigma _{x}^{D}+2\sigma _{y}^{R}\sigma _{y}^{D}=2\left(
G_{1,x}G_{2,y}+G_{1,y}G_{2,x}\right) \sin \theta \cos \theta \left(
1+r^{4}W^{2}\sin ^{2}\theta \cos ^{2}\theta \right) .
\end{equation*}%
The nominator is
\begin{eqnarray*}
&&\left( \sigma _{x}^{R}+\sigma _{x}^{D}\right) d\left( \sigma
_{y}^{R}+\sigma _{y}^{D}\right) -\left( \sigma _{y}^{R}+\sigma
_{y}^{D}\right) d\left( \sigma _{x}^{R}+\sigma _{x}^{D}\right)  \\
&=&\left( \sigma _{x}^{R}d\sigma _{y}^{R}-\sigma _{y}^{R}d\sigma
_{x}^{R}+\sigma _{x}^{D}d\sigma _{y}^{D}-\sigma _{y}^{D}d\sigma
_{x}^{D}\right) +\left( \sigma _{x}^{R}d\sigma _{y}^{D}+\sigma
_{x}^{D}d\sigma _{y}^{R}-\sigma _{y}^{R}d\sigma _{x}^{D}-\sigma
_{y}^{D}d\sigma _{x}^{R}\right) ,
\end{eqnarray*}%
and%
\begin{eqnarray*}
\sigma _{x}^{R}d\sigma _{y}^{D} &=&\left( G_{1,x}\cos \theta
+r^{2}G_{1,y}W\cos \theta \sin ^{2}\theta \right) \left( -G_{2,x}\sin \theta
+r^{2}G_{2,y}W\sin ^{3}\theta -2r^{2}G_{2,y}W\cos ^{2}\theta \sin \theta
\right) , \\
\sigma _{x}^{D}d\sigma _{y}^{R} &=&\left( G_{2,y}\sin \theta
+r^{2}G_{2,x}W\sin \theta \cos ^{2}\theta \right) \left( G_{1,y}\cos \theta
-r^{2}G_{1,x}W\cos ^{3}\theta +2r^{2}G_{1,x}W\cos \theta \sin ^{2}\theta
\right) , \\
\sigma _{y}^{R}d\sigma _{x}^{D} &=&\left( G_{1,y}\sin \theta
-r^{2}G_{1,x}W\cos ^{2}\theta \sin \theta \right) \left( G_{2,y}\cos \theta
+r^{2}G_{2,x}W\cos ^{3}\theta -2r^{2}G_{2,x}W\sin ^{2}\theta \cos \theta
\right) , \\
\sigma _{y}^{D}d\sigma _{x}^{R} &=&\left( G_{2,x}\cos \theta
-r^{2}G_{2,y}W\cos \theta \sin ^{2}\theta \right) \left( -G_{1,x}\sin \theta
-r^{2}G_{1,y}W\sin ^{3}\theta +2r^{2}G_{1,y}W\cos ^{2}\theta \sin \theta
\right) .
\end{eqnarray*}%
Then we have
\begin{equation*}
\sigma _{x}^{R}d\sigma _{y}^{D}+\sigma _{x}^{D}d\sigma _{y}^{R}-\sigma
_{y}^{R}d\sigma _{x}^{D}-\sigma _{y}^{D}d\sigma _{x}^{R}=-\frac{1}{2}%
r^{2}W\left( \sin 4\theta \right) \left(
G_{2,x}G_{1,y}+G_{2,y}G_{1,x}\right) .
\end{equation*}%
It is straightforward to obtain
\begin{equation*}
q=\frac{1}{2\pi }\int_{0}^{2\pi }d\theta \frac{\left(
G_{1,x}G_{1,y}-G_{2,y}G_{2,x}\right) }{\left( G_{1,x}\cos \theta
+G_{2,y}\sin \theta \right) ^{2}+\left( G_{1,y}\sin \theta +G_{2,x}\cos
\theta \right) ^{2}}-\frac{1}{2\pi }\int_{0}^{2\pi }d\theta \frac{r^{2}W\cos
\left( 2\theta \right) }{\left( 1+r^{4}W^{2}\cos ^{2}\theta \sin ^{2}\theta
\right) }.
\end{equation*}%
The elliptic effect gives the same integral in any case of calculating the
topological charge, which is zero. The topological charge is
\begin{equation*}
q=\frac{1}{\pi }\int_{0}^{\pi }\frac{G_{1,x}G_{1,y}-G_{2,x}G_{2,y}}{A+B\cos t%
}dt=\frac{G_{1,x}G_{1,y}-G_{2,x}G_{2,y}}{\sqrt{\left(
G_{1,x}G_{1,y}-G_{2,x}G_{2,y}\right) ^{2}}},
\end{equation*}%
where%
\begin{eqnarray*}
A &=&\frac{1}{2}G_{1,x}^{2}+\frac{1}{2}G_{2,x}^{2}+\frac{1}{2}G_{1,y}^{2}+%
\frac{1}{2}G_{2,y}^{2}, \\
B &=&\sqrt{\frac{1}{4}\left( \allowbreak
G_{1,x}^{2}+G_{2,x}^{2}-G_{1,y}^{2}-G_{2,y}^{2}\right) ^{2}+\left(
G_{1,x}G_{2,y}+G_{2,x}\allowbreak G_{1,y}\right) ^{2}} \\
A^{2}-B^{2} &=&\left( G_{1,x}G_{1,y}-G_{2,x}G_{2,y}\right) ^{2}.
\end{eqnarray*}%
Finally, we obtain that the topological charge is equivalent to Eq. (\ref{tc}%
).

\end{widetext}

\end{document}